\pdfoutput=1
\documentclass{article}
\usepackage{arxiv}
\usepackage[utf8]{inputenc}
\usepackage[T1]{fontenc}
\usepackage{amsmath, amsfonts, amssymb, amsthm}
\usepackage{booktabs}
\usepackage{hyperref}
\usepackage{algorithm}
\usepackage{algpseudocode}
\usepackage{bm}
\usepackage{multirow}
\usepackage{caption}
\newtheorem{remark}{Remark}
\newtheorem{proposition}{Proposition}
\newtheorem{theorem}{Theorem}

\newcommand{\bgamma}{\bm{\gamma}}
\newcommand{\btheta}{\bm{\theta}}
\newcommand{\bx}{\bm{x}}
\newcommand{\X}{\mathbf{X}}
\newcommand{\y}{\bm{y}}

\newcommand{\W}{\mathbf{W}}

\title{Regularized parametric modal regression\\
for positive responses with\\
high-dimensional covariates}

\author{
  Christian E.\ Galarza \\
  Escuela Superior Polit\'ecnica del Litoral\\
  Guayaquil, Ecuador \\
  \texttt{chedgala@espol.edu.ec}
  \And
  V\'ictor H.\ Lachos \\
  University of Connecticut \\
  Storrs, CT, USA \\
  \texttt{hlachos@uconn.edu}
  \And
  Dashun Liu \\
  University of Connecticut \\
  Storrs, CT, USA \\
  \texttt{dashun.liu@uconn.edu}
}

\begin{document}
\maketitle

\begin{abstract}
Modal regression models the conditional mode rather than the conditional mean, which is attractive for positive and right-skewed responses whose mean is distorted by the upper tail. We propose a regularized version of parametric modal regression for positive responses with high-dimensional covariates, in which the density parameters of the Gamma, Weibull and Lognormal distributions are written as functions of the conditional mode and the log-likelihood is penalized by the Lasso, adaptive Lasso or elastic-net penalties. The estimates are computed by a proximal Newton algorithm, equivalently an iteratively reweighted penalized least-squares scheme, whose inner problem is solved exactly by the coordinate-descent routines of \texttt{glmnet}, and the regularization parameter is selected by the extended Bayesian information criterion. We show that the Lognormal estimator coincides with the Gaussian Lasso applied to the log-transformed response, that under constant dispersion the conditional mean and the conditional mode have the same slopes, and that the estimation error attains the rate $\mathcal{O}(\sqrt{s}\lambda_n)$ with no smoothing bias. Simulation studies with up to $p = 1000$ covariates show that the method improves on the Lasso applied to the log-transformed response for Gamma and Weibull responses, and two applications illustrate its use. The method is implemented in the \texttt{R} package \texttt{ModalNet}.
\end{abstract}


\section{Introduction}
\label{sec:intro}

Classical generalized linear models (GLMs; \cite{mccullagh1989}) target
the conditional mean $\mathbb{E}[Y_i \mid \bx_i]$ as the primary measure
of location. For positive and right-skewed response distributions---such as
those arising in survival analysis, reliability engineering, insurance
claims, and environmental monitoring---the conditional mean is sensitive to
extreme observations in the upper tail and may not correspond to any
high-density region of the response. In such contexts, the conditional
\emph{mode} $M_i = \arg\max_{y} f(y \mid \bx_i)$ is a more
interpretable characterization of the ``most likely'' outcome.
As shown in Section~\ref{sec:mode_mean}, under a logarithmic link with
constant dispersion this distinction concerns the estimand and the fitted
intercept, whereas the slope coefficients coincide with those of the mean
model.

Modal regression---modeling $M_i$ as a function of covariates---has
attracted increasing attention, with contributions spanning nonparametric
methods \cite{yao2014,chen2016} and Bayesian approaches \cite{zhou2025}.
Galarza and Lachos \cite{galarza2025modal} introduced a parametric framework
for continuous positive distributions, in which the density parameters are
written as functions of the conditional mode. Variable selection has been
studied for kernel-based modal regression \cite{wang2020robust} and in a
Bayesian framework \cite{zhou2025}, but not for this parametric formulation.

In parallel, penalized regression methods have become indispensable
tools for high-dimensional inference. The Lasso \cite{tibshirani1996}
performs simultaneous coefficient estimation and variable selection by
imposing an $L_1$ penalty on the regression coefficients. Its properties
for mean regression are well studied \cite{buhlmann2011,freijeiro2022}, and
efficient coordinate-descent solvers are widely available \cite{friedman2010}.

A common strategy for extending these solvers beyond the linear model is
the penalized EM algorithm \cite{dempster1977,green1990}, whose M-step is an
ordinary penalized regression once the missing information has been imputed.
It has been used for censored responses and, with the random effects as
missing data, for high-dimensional linear mixed-effects models
\cite{oliveira2025,schelldorfer2011}, in both cases delegating the M-step to
\texttt{glmnet} \cite{friedman2010}.

In this paper we propose a regularized version of the parametric modal
regression model of \cite{galarza2025modal} for positive responses with
high-dimensional covariates, under the Gamma, Weibull and Lognormal
distributions. The penalized modal log-likelihood is non-differentiable,
through the penalty, and its curvature in the regression coefficients is not
constant, through the exponential link on the mode. Instead of replacing the
$L_1$ norm by a smooth surrogate, which biases the solution and yields no exact
zeros, at each iteration we replace the log-likelihood by a quadratic
approximation in the linear predictor, so that the update is a penalized
weighted least-squares problem solved exactly by \texttt{glmnet}, as in the
M-step of the penalized EM algorithms above. The regularization parameter is
selected along a warm-started path by the extended Bayesian information
criterion \cite{chen2008}. The method is implemented in the \texttt{R} package
\texttt{ModalNet} \cite{modalnet}, available from CRAN.

The paper is organized as follows. Section~\ref{sec:prelim} reviews the Lasso
and the penalized EM algorithm. Section~\ref{sec:model} presents the model,
the algorithm and its properties. Sections~\ref{sec:simulation}
and~\ref{sec:application} report simulation studies and two applications, and
Section~\ref{sec:conclusion} concludes.

\section{Preliminaries}
\label{sec:prelim}

We begin by defining the notation and presenting some basic concepts used
throughout the development of our methodology. As is usual in probability
theory and its applications, we denote a random variable by an upper-case
letter and its realization by the corresponding lower-case letter, and we use
boldface letters for vectors and matrices. $\mathrm{N}(\mu, \sigma^2)$ denotes
the normal distribution with mean $\mu$ and variance $\sigma^2$, and
$\mathrm{U}(a,b)$ the uniform distribution on $(a,b)$. $\mathbf{A}^\top$
denotes the transpose of a matrix $\mathbf{A}$, and \textit{iid} stands for
independent and identically distributed.

\subsection{Lasso and regularization}
\label{sec:lasso}

First, consider the standard linear regression model
\begin{equation}
  Y_i = \bx_i^\top \bm{\beta} + \varepsilon_i, \qquad i = 1, \ldots, n,
  \label{eq:linear_model}
\end{equation}
where $Y_i$ is a continuous outcome, $\bx_i$ is a $p \times 1$ vector of
fixed predictors and $\bm{\beta} = (\beta_1, \ldots, \beta_p)^\top$ is a
$p \times 1$ vector of unknown regression parameters. The Lasso estimator of
the parameters in model \eqref{eq:linear_model} is
\begin{equation}
  \hat{\bm{\beta}}(\lambda) = \arg\min_{\bm{\beta}}
    \left\{ \sum_{i=1}^n \left(Y_i - \bx_i^\top\bm{\beta}\right)^2
      + \lambda\, \Psi_p(\bm{\beta}) \right\},
  \label{eq:lasso}
\end{equation}
where $\Psi_p(\bm{\beta}) = \|\bm{\beta}\|_1 = \sum_{j=1}^p |\beta_j|$ ($L_1$
norm) is the Lasso penalty on the parameter size and $\lambda \geq 0$ controls
the amount of regularization. When $\lambda$ is large enough, all coefficients
are shrunk exactly to zero; conversely, $\lambda = 0$ recovers the unpenalized
least-squares estimate.

When $\Psi_p(\bm{\beta}) = \|\bm{\beta}\|_2^2 = \sum_{j=1}^p \beta_j^2$ ($L_2$
norm) we obtain the ridge penalty, which reduces model complexity but does not
force any coefficient to be exactly zero and is therefore unsuitable for
variable selection. The elastic net \cite{zou2005} combines both penalties,
\begin{equation}
  \Psi_p^{\alpha}(\bm{\beta}) = \alpha\|\bm{\beta}\|_1
    + (1-\alpha)\|\bm{\beta}\|_2^2 ,
\end{equation}
so that $\alpha = 1$ recovers the Lasso and $\alpha \to 0$ approaches ridge
regression. A further refinement is the adaptive Lasso \cite{zou2006}, which
replaces $\Psi_p$ by a weighted $L_1$ norm $\sum_j w_j|\beta_j|$ with data-driven
weights, thereby reducing the shrinkage bias on large coefficients. For
simplicity we write $\Psi_p(\bm{\beta})$ for a generic penalty function
throughout.

In \texttt{R}, an efficient implementation is available in the package
\texttt{glmnet} \cite{friedman2010}, which fits generalized linear models via
penalized maximum likelihood (PML) using cyclical coordinate descent. The
algorithm is designed for high-speed computation, exploits sparsity in the
input matrix $\X$, and accommodates observation weights. Our proposal takes
advantage of this available software, since each iteration of the
algorithm of Section~\ref{sec:emlasso} is a penalized weighted least-squares
problem of this form.

\subsection{The penalized EM algorithm for PML}
\label{sec:pml}

Assume that the data $\{y_i, \bx_i;\, i = 1, \ldots, n\}$ are collected
independently and that $\btheta = (\bgamma^\top, \bm{\rho}^\top)^\top$
collects the model parameters, where $\bgamma$ contains the regression
coefficients subject to penalization and $\bm{\rho}$ the remaining
(nuisance) parameters. Let $\ell(\btheta) = \sum_{i=1}^n \ell_i(\btheta)$
denote the log-likelihood. A form of the penalized log-likelihood is
\begin{equation}
  \ell_p(\btheta) = \ell(\bgamma, \bm{\rho}) - \lambda\, \Psi_p(\bgamma).
  \label{eq:penlik}
\end{equation}
Maximizing \eqref{eq:penlik} rather than $\ell(\btheta)$ yields estimates in
which some components of $\bgamma$ are exactly zero, so that variable
selection and parameter estimation are performed in a single step.

The EM algorithm \cite{dempster1977} is a popular iterative procedure for
maximum likelihood estimation in models with incomplete data, valued for its
monotone convergence and ease of implementation. Following the penalized EM
algorithm of \cite{green1990}, suppose the observed data are augmented by a
latent quantity, and let $\ell_c^p(\btheta)$ denote the resulting penalized
complete-data log-likelihood. The algorithm iterates the following two steps.
\begin{itemize}
  \item[$\bullet$] {\bf E-step.} Compute the conditional expectation of
    $\ell_c^p(\btheta)$ given the observed data, evaluated at the current
    estimate $\btheta^{(k)}$, yielding the penalized $Q^p$-function
    \[
      Q^p(\btheta \mid \btheta^{(k)})
        = \mathbb{E}_{\btheta^{(k)}}\!\left[\ell_c(\btheta) \mid \y\right]
          - \lambda\, \Psi_p(\bgamma).
    \]
    In the applications of interest this function separates as
    \begin{equation}
      Q^p(\btheta \mid \btheta^{(k)})
        = f(\bm{\rho}, \btheta^{(k)})\, Q_1^p(\bgamma, \lambda \mid \btheta^{(k)})
          + Q_2(\bm{\rho} \mid \btheta^{(k)}),
      \label{eq:Qsplit}
    \end{equation}
    where $f(\cdot)$ depends only on $\bm{\rho}$, so that maximizing
    $Q^p$ with respect to $\bgamma$ is equivalent to maximizing
    $Q_1^p(\bgamma, \lambda \mid \btheta^{(k)})$ for fixed $\lambda$.
  \item[$\bullet$] {\bf M-step.} Maximize $Q^p(\btheta \mid \btheta^{(k)})$
    with respect to $\btheta$. From \eqref{eq:Qsplit}, and keeping $\lambda$
    fixed, the parameters are updated by
    \begin{align*}
      \bgamma^{(k+1)}(\lambda) &= \arg\min_{\bgamma}
        \left\{ Q_1^p(\bgamma, \lambda \mid \btheta^{(k)}) \right\}, \\
      \bm{\rho}^{(k+1)} &= \arg\min_{\bm{\rho}}
        \left\{ Q^p(\bgamma^{(k+1)}, \bm{\rho} \mid \btheta^{(k)}) \right\}.
    \end{align*}
\end{itemize}
The algorithm is terminated when the relative distance between two successive
evaluations of the penalized log-likelihood \eqref{eq:penlik} is smaller than
a tolerance, that is,
$|\ell_p(\btheta^{(k+1)})/\ell_p(\btheta^{(k)}) - 1| < \epsilon$, for example
$\epsilon = 10^{-6}$.

The key structural requirement is that $Q_1^p$ be a penalized least-squares
objective, since it is precisely this form that the coordinate-descent solvers
of Section~\ref{sec:lasso} are built to handle. Section~\ref{sec:emlasso}
shows that the modal regression model admits such a representation.

\section{The Penalized Modal Regression Model}
\label{sec:model}

\subsection{Model and penalized likelihood}
\label{sec:framework}

We briefly recall the modal parameterization developed in
\cite{galarza2025modal}. Let $Y_i$ be a continuous positive random variable
whose density $f(y_i \mid \btheta)$ admits a unique interior mode. The
conditional mode satisfies the first-order stationarity condition
\begin{equation}
  \left.\frac{\partial \log f(y_i \mid \btheta)}{\partial y_i}
  \right|_{y_i = M_i} = 0.
  \label{eq:foc}
\end{equation}
Solving \eqref{eq:foc} for one distributional parameter as a function of
$M_i$ and the remaining parameters yields the \emph{mode reparameterization}:
a density in which the mode $M_i$ replaces one of the original parameters.

Let $\bx_i = (1, x_{i1}, \ldots, x_{ip})^\top \in \mathbb{R}^{p+1}$
denote the covariate vector (including an intercept) for observation $i$,
$i = 1, \ldots, n$. The modal regression model specifies
\begin{equation}
  g(M_i) = \bx_i^\top \bgamma,
  \label{eq:link}
\end{equation}
where $g(\cdot)$ is a monotone, differentiable link function and
$\bgamma = (\gamma_0, \gamma_1, \ldots, \gamma_p)^\top \in
\mathbb{R}^{p+1}$ is the vector of regression coefficients. Throughout this
work we adopt the logarithmic link $g(M_i) = \log(M_i)$, so that
$M_i = \exp(\bx_i^\top \bgamma)$. The full parameter vector is
$\btheta = (\bgamma^\top, \phi)^\top$, where $\phi > 0$ is a global
dispersion parameter, playing the role of the nuisance component
$\bm{\rho}$ of Section~\ref{sec:pml}. We write
$\eta_i = \bx_i^\top\bgamma$ for the linear predictor.

The mode reparameterizations for the three families considered here are as
follows.

\paragraph{Gamma distribution.}
For $Y_i \sim \mathrm{Gamma}(\alpha, \beta_i)$ with shape $\alpha > 1$ and
scale $\beta_i$, the mode is $M_i = (\alpha - 1)\beta_i$. Setting
$\phi = 1/(\alpha - 1)$:
\begin{equation}
  \beta_i = M_i \phi, \quad \alpha = 1/\phi + 1.
  \label{eq:gamma_map}
\end{equation}

\paragraph{Weibull distribution.}
For $Y_i \sim \mathrm{Weibull}(k^{\ast}, \lambda_i)$ with shape
$k^{\ast} > 1$ and scale $\lambda_i$, the mode is
$M_i = \lambda_i [(k^{\ast}-1)/k^{\ast}]^{1/k^{\ast}}$. Setting
$k^{\ast} = \phi + 1.01$:
\begin{equation}
  \lambda_i = M_i \left(\frac{k^{\ast}}{k^{\ast}-1}\right)^{1/k^{\ast}}.
  \label{eq:weibull_map}
\end{equation}
The superscript distinguishes the Weibull shape from the iteration index
$k$ used in Section~\ref{sec:emlasso}.

\paragraph{Lognormal distribution.}
For $Y_i \sim \mathrm{Lognormal}(\mu_i, \sigma^2)$, the mode is
$M_i = \exp(\mu_i - \sigma^2)$. Setting $\sigma^2 = \phi$:
\begin{equation}
  \mu_i = \log(M_i) + \phi.
  \label{eq:lognormal_map}
\end{equation}

The log-likelihood for the modal regression model, given
$\{(y_i, \bx_i)\}_{i=1}^n$, takes the general form
\begin{equation}
  \ell(\btheta) = \ell(\bgamma, \phi) = \sum_{i=1}^{n}
  \log f\!\left(y_i \mid M_i(\bgamma), \phi\right),
  \label{eq:loglik}
\end{equation}
where $M_i(\bgamma) = \exp(\bx_i^\top \bgamma)$ and
$f(\cdot \mid M_i, \phi)$ denotes the density under the chosen family
after substituting the mode reparameterization. Following
\eqref{eq:penlik}, the penalized log-likelihood is
\begin{equation}
  \ell_p(\btheta) = \ell(\bgamma, \phi) - \lambda\, \Psi_p(\bgamma),
  \label{eq:penloglik_modal}
\end{equation}
where the penalty is applied to $\gamma_1, \ldots, \gamma_p$ and excludes
the intercept $\gamma_0$.

The Lasso, adaptive Lasso and elastic-net penalties of
Section~\ref{sec:lasso} differ only in the penalty passed to the inner solver
of Section~\ref{sec:emlasso}, so that they need no further derivation; the
ridge penalty yields no exact zeros and is not considered here.

\subsection{The proximal Newton algorithm}
\label{sec:emlasso}

Objective \eqref{eq:penloglik_modal} is non-differentiable, through the
penalty, and non-quadratic in $\bgamma$, through the exponential link. We
replace the log-likelihood at each iteration by a quadratic approximation in
the linear predictor and keep the penalty exact, so that the problem takes the
form \eqref{eq:Qsplit} of Section~\ref{sec:pml}.

\begin{remark}[Relation to penalized EM algorithms]
\label{rem:not_em}
The iteration is a proximal Newton method \cite{lee2014}, or equivalently
an iteratively reweighted penalized least-squares (IRPLS) scheme. It is not an
EM algorithm, since the model has no latent variable; it shares with the
penalized EM algorithms of Section~\ref{sec:pml} only the form of the inner
problem, a weighted sum of squares plus a penalty.
\end{remark}

{\bf Quadratic-approximation step.} Fix the dispersion $\phi$ and let $\bgamma^{(k)}$ denote the
current iterate, with $\eta_i^{(k)} = \bx_i^\top\bgamma^{(k)}$. Because the
log-likelihood \eqref{eq:loglik} depends on $\bgamma$ only through the linear
predictors, each contribution $\ell_i(\eta_i)$ may be expanded to second order
about $\eta_i^{(k)}$. Writing
\begin{equation}
  s_i^{(k)} = \left.\frac{\partial \ell_i}{\partial \eta_i}\right|_{\eta_i^{(k)}},
  \qquad
  w_i^{(k)} = \left.-\frac{\partial^2 \ell_i}{\partial \eta_i^2}\right|_{\eta_i^{(k)}},
  \label{eq:scorecurv}
\end{equation}
and completing the square gives
\begin{equation}
  \ell_i(\eta_i) \;\simeq\; \text{const} \;-\; \frac{w_i^{(k)}}{2}
  \left(z_i^{(k)} - \eta_i\right)^2,
  \qquad
  z_i^{(k)} = \eta_i^{(k)} + \frac{s_i^{(k)}}{w_i^{(k)}},
  \label{eq:surrogate}
\end{equation}
where $z_i^{(k)}$ is the \emph{pseudo-response} and $w_i^{(k)}$ the
\emph{working weight}, that is, the observed information in $\eta_i$.
For the three families $w_i^{(k)} > 0$ whenever $y_i > 0$, so that the update
is a convex problem. Since \eqref{eq:surrogate} is only a local
approximation, a full step may increase the penalized objective, and we add a
backtracking line search: given the solution $\tilde{\bgamma}$ of
\eqref{eq:mstep}, we set
$\bgamma^{(k+1)} = \bgamma^{(k)} + \alpha^{(k)}(\tilde{\bgamma} - \bgamma^{(k)})$,
with $\alpha^{(k)}$ the largest element of $\{1, \tfrac12, \tfrac14, \ldots\}$
that does not increase the objective, which makes the iteration monotone
\cite{lee2014}.

Carrying out the differentiation in \eqref{eq:scorecurv} for each family gives
the following closed-form quantities, where $M_i^{(k)} = \exp(\eta_i^{(k)})$.

\paragraph{Gamma.} With $\alpha = 1/\phi + 1$ and $\beta_i = \phi M_i$ we have
$s_i = y_i/(\phi M_i) - \alpha$ and $w_i = y_i/(\phi M_i)$, so that
\begin{equation}
  w_i^{(k)} = \frac{y_i}{\phi\, M_i^{(k)}},
  \qquad
  z_i^{(k)} = \eta_i^{(k)} + 1 - \frac{(1+\phi)\,M_i^{(k)}}{y_i}.
  \label{eq:gamma_em}
\end{equation}

\paragraph{Weibull.} With $k^{\ast} = \phi + 1.01$ and
$\lambda_i = M_i\{k^{\ast}/(k^{\ast}-1)\}^{1/k^{\ast}}$ we obtain
$s_i = k^{\ast}\{(y_i/\lambda_i)^{k^{\ast}} - 1\}$ and
$w_i = k^{\ast 2}(y_i/\lambda_i)^{k^{\ast}}$, so that
\begin{equation}
  w_i^{(k)} = k^{\ast 2}\!\left(\frac{y_i}{\lambda_i^{(k)}}\right)^{\!k^{\ast}},
  \qquad
  z_i^{(k)} = \eta_i^{(k)}
    + \frac{1}{k^{\ast}}\!\left[1 - \left(\frac{\lambda_i^{(k)}}{y_i}\right)^{\!k^{\ast}}\right].
  \label{eq:weibull_em}
\end{equation}

\paragraph{Lognormal.} With $\mu_i = \eta_i + \phi$ and $\sigma^2 = \phi$ we
have $s_i = (\log y_i - \eta_i - \phi)/\phi$ and $w_i = 1/\phi$, so that
\begin{equation}
  w_i = \frac{1}{\phi} \quad (\text{constant in } i),
  \qquad
  z_i = \log y_i - \phi.
  \label{eq:lognormal_em}
\end{equation}
Here the surrogate is \emph{exact} rather than approximate, since the
Lognormal log-likelihood is already quadratic in $\eta_i$; neither $w_i$ nor
$z_i$ depends on the current iterate.

{\bf Update step.} Substituting \eqref{eq:surrogate} into
\eqref{eq:penloglik_modal}, the update for $\bgamma$ is the penalized weighted
least-squares problem
\begin{equation}
  \bgamma^{(k+1)}
  = \arg\min_{\bgamma}
    \left\{ \frac{1}{2}\sum_{i=1}^n w_i^{(k)}
      \left(z_i^{(k)} - \bx_i^\top\bgamma\right)^2
      + \lambda\, \Psi_p(\bgamma) \right\},
  \label{eq:mstep}
\end{equation}
which is convex and is solved exactly by the coordinate descent of
\texttt{glmnet} \cite{friedman2010} through its observation weights. Its
solution contains exact zeros, so that no thresholding is needed for variable
selection.

Given the slopes, the intercept and the dispersion, which are not penalized
and are strongly coupled in the mode parameterization (for instance,
$\beta_i = \phi M_i$ for the Gamma family), are updated jointly by maximizing
$\ell(\bgamma, \phi)$ over $(\gamma_0, \phi)$. For the Lognormal family, with
$r_i = \log y_i - \sum_{j=1}^p x_{ij}\gamma_j$ and $\bar r = n^{-1}\sum_i r_i$,
\begin{equation}
  \phi^{(k+1)} = \frac{1}{n}\sum_{i=1}^n (r_i - \bar r)^2,
  \qquad
  \gamma_0^{(k+1)} = \bar r - \phi^{(k+1)}.
  \label{eq:phi_lognormal}
\end{equation}
and for the Gamma and Weibull families $(\gamma_0, \log\phi)$ is updated by a
two-dimensional Nelder--Mead search, with $\phi \in [10^{-4}, 10^{4}]$. The
iteration stops when the relative change of the penalized objective, or the
change of the iterates, is below $\epsilon = 10^{-6}$, with at most $200$
iterations per value of $\lambda$. If $\phi$ stays at an endpoint of its range
during ten consecutive iterations, the fit is stopped and flagged as not
converged; for the Gamma ($\phi \to \infty$) and Weibull ($\phi \to 0$)
families this endpoint is the exponential limit, at which the mode tends to
zero.

\subsection{Exact equivalence in the Lognormal case}

The constancy of the working weight in \eqref{eq:lognormal_em} has the
following consequence.

\begin{proposition}[Lognormal equivalence]
\label{prop:lognormal}
Consider the Lognormal modal regression model with dispersion $\phi$ held
fixed and $\Psi_p$ the Lasso penalty. Let $\hat{\bgamma}^{\mathrm{modal}}(\lambda)$
maximize \eqref{eq:penloglik_modal}, and let
$\hat{\bgamma}^{\mathrm{Gauss}}(\lambda')$ minimize
$\tfrac{1}{2}\|\log\y - \X\bgamma\|_2^2 + \lambda'\|\bgamma_{-0}\|_1$. Then for
every $\lambda > 0$,
\begin{equation}
  \hat\gamma_j^{\mathrm{modal}}(\lambda) = \hat\gamma_j^{\mathrm{Gauss}}(\phi\lambda),
  \quad j = 1,\ldots,p,
  \qquad
  \hat\gamma_0^{\mathrm{Gauss}}(\phi\lambda) - \hat\gamma_0^{\mathrm{modal}}(\lambda) = \phi .
  \label{eq:lognormal_equiv}
\end{equation}
\end{proposition}

\begin{proof}
Since the Lognormal log-likelihood is quadratic in $\eta_i$,
\eqref{eq:penloglik_modal} equals, up to a constant,
$-(2\phi)^{-1}\sum_i (z_i - \bx_i^\top\bgamma)^2 - \lambda\|\bgamma_{-0}\|_1$,
with $z_i = \log y_i - \phi$. Multiplying by $-\phi$ gives the Gaussian Lasso
objective with response $\log\y - \phi\bm{1}$ and penalty $\phi\lambda$. Since
the intercept is not penalized, replacing this response by $\log\y$ shifts the
fitted intercept by $\phi$ and leaves the slopes unchanged.
\end{proof}

Hence, for lognormal responses, the Gaussian Lasso applied to $\log\y$ is
the modal Lasso up to the interpretation of the intercept. The factor $\phi$
disappears in the implementation, since \texttt{glmnet} rescales the weights
$w_i = 1/\phi$ to one, so that both fits coincide at equal $\lambda$, as
observed in Section~\ref{sec:application}. The argument relies on the
constancy of $w_i$, which fails for the Gamma and Weibull families, and the
simulations of Section~\ref{sec:simulation} show that the two procedures then
differ.

\subsection{Relation between the conditional mode and the conditional mean}
\label{sec:mode_mean}

The next result relates the conditional mode to the conditional mean.

\begin{proposition}[Mode--mean proportionality]
\label{prop:modemean}
Let $M_i = \exp(\bx_i^\top\bgamma)$ be the conditional mode under any of the
three families of Section~\ref{sec:framework}, with $\phi$ fixed. Then
$\mathbb{E}[Y_i \mid \bx_i] = c(\phi)\, M_i$, where
\begin{equation*}
  c(\phi) =
  \begin{cases}
    1 + \phi, & \text{Gamma},\\[2pt]
    \left(\dfrac{k^{\ast}}{k^{\ast}-1}\right)^{1/k^{\ast}}
      \Gamma\!\left(1 + 1/k^{\ast}\right), \quad k^{\ast} = \phi + 1.01,
      & \text{Weibull},\\[6pt]
    \exp(3\phi/2), & \text{Lognormal},
  \end{cases}
\end{equation*}
so that $\log \mathbb{E}[Y_i \mid \bx_i] = \log c(\phi) + \bx_i^\top\bgamma$.
\end{proposition}

\begin{proof}
From \eqref{eq:gamma_map}, the Gamma mean is
$\alpha\beta_i = (1+\phi)M_i$. From \eqref{eq:weibull_map}, the Weibull mean
is $\lambda_i\Gamma(1 + 1/k^{\ast})$, with $\lambda_i$ proportional to $M_i$.
From \eqref{eq:lognormal_map}, the Lognormal mean is
$\exp(\mu_i + \phi/2) = \exp(3\phi/2) M_i$.
\end{proof}

Hence, under constant dispersion, modal and mean regression have the same
slopes and the same active set, and differ only in the intercept, since
$\exp(\hat\gamma_0)$ is the most likely response at $\bx = \bm{0}$ instead of
the expected one. If the dispersion depends on the covariates, $c(\phi_i)$
varies with $\bx_i$ and the slopes no longer coincide.

\subsection{Tuning and path following}
\label{sec:tuning}

Since the BIC tends to select too many noise variables in high
dimensions, we select $\lambda$ by minimizing the extended Bayesian
information criterion \cite{chen2008},
\begin{equation}
  \mathrm{EBIC}(\lambda) = -2\,\ell(\hat{\bgamma}_\lambda, \hat{\phi}_\lambda)
  + \mathrm{df}(\lambda)\,\log n + 2\,\gamma_{\mathrm{EBIC}}\,\mathrm{df}(\lambda)\log p,
  \label{eq:ebic}
\end{equation}
where $\mathrm{df}(\lambda) = |\hat{\mathcal{S}}_\lambda| + 2$ counts the
active coefficients plus the intercept and the dispersion parameter, and
$\gamma_{\mathrm{EBIC}} \in [0,1]$ governs the additional penalty on the
model space, with $\gamma_{\mathrm{EBIC}} = 0.5$ as default.

As is standard for penalized GLMs, the criterion is evaluated along a
path: a decreasing grid
$\lambda_1 > \cdots > \lambda_L$ is constructed once from the initial
pseudo-response, the algorithm is run \emph{to convergence} at each grid
point, warm-started at the solution for the previous and larger value, and
\eqref{eq:ebic} is evaluated across the resulting fits. By default $L = 50$ and
$\lambda_L/\lambda_1$ equals $0.01$ when $n > p$ and $0.05$ otherwise, as in
\texttt{glmnet}. The path is stopped once more than $d_{\max}$ covariates are
active, since near saturation the likelihood grows without bound and the
criterion \eqref{eq:ebic} is no longer informative; the default is
$d_{\max} = \min\{p, \max(5, \lfloor 0.8n \rfloor)\}$. The null model, with
only the intercept and the dispersion, is always among the candidates, since
the first point of the grid need not be empty.
Algorithm~\ref{alg:emlasso} summarizes the procedure.

\begin{algorithm}[ht]
\caption{Proximal Newton algorithm for penalized modal regression}
\label{alg:emlasso}
\begin{algorithmic}[1]
\Require Response $\y$, design $\X$, family, grid length $L$, $\gamma_{\mathrm{EBIC}}$, tolerance $\epsilon$, maximum model size $d_{\max}$.
\State \textbf{Initialize:} $\bgamma^{(0)}$ from a lightly penalized Lasso fit of $\log \y$ on $\X$ (at $0.1$ times the largest penalty of its path); $\phi^{(0)} \leftarrow \max\{\mathrm{Var}(\log \y - \X\bgamma^{(0)}),\, 0.05\}$; then update $(\gamma_0^{(0)}, \phi^{(0)})$ jointly as in step~9.
\State Form the pair $(z_i, w_i)$ at $(\bgamma^{(0)}, \phi^{(0)})$ and build the grid $\lambda_1 > \cdots > \lambda_L$ from it.
\For{$\ell = 1, \ldots, L$}
  \State Warm start at the solution for $\lambda_{\ell-1}$.
  \Repeat
    \State \textbf{Approximation:} compute $w_i^{(k)}, z_i^{(k)}$ from \eqref{eq:gamma_em}, \eqref{eq:weibull_em} or \eqref{eq:lognormal_em}.
    \State \textbf{Update ($\bgamma$):} solve \eqref{eq:mstep} at $\lambda_\ell$ by coordinate descent to obtain the trial point $\tilde{\bgamma}$.
    \State \textbf{Line search:} set $\bgamma^{(k+1)} \leftarrow \bgamma^{(k)} + \alpha^{(k)}(\tilde{\bgamma} - \bgamma^{(k)})$ with $\alpha^{(k)} \in \{1, \tfrac12, \tfrac14, \ldots\}$ the largest value for which the penalized objective does not increase.
    \State \textbf{Update ($\gamma_0, \phi$):} maximize $\ell$ over $(\gamma_0, \phi)$ with the slopes fixed, by \eqref{eq:phi_lognormal} or by a two-dimensional search.
  \Until{the relative change of $\ell_p$ is below $\epsilon$, or the iterates no longer change.}
  \State Record the fit and evaluate $\mathrm{EBIC}(\lambda_\ell)$ from \eqref{eq:ebic}.
  \State \textbf{if} more than $d_{\max}$ covariates are active \textbf{then} discard this fit and stop the path.
\EndFor
\Ensure $\hat\lambda = \arg\min_\ell \mathrm{EBIC}(\lambda_\ell)$ over the path and the null model, and the associated $\hat{\bgamma}, \hat\phi, \hat{\mathcal{S}}$.
\end{algorithmic}
\end{algorithm}

The initialization in step~1 uses a lightly penalized Lasso fit of
$\log\y$, since the least-squares fit is not defined when $p + 1 > n$, and the
linear predictor is clamped to $[-30, 30]$ to prevent overflow in
$\exp(\eta_i)$.

\subsection{Theoretical consistency of the penalized estimator}
\label{sec:theory}

With $\phi$ fixed, let
$\mathcal{L}_n(\bgamma) = -\frac{1}{n}\sum_{i=1}^n \log f(y_i \mid M_i(\bgamma), \phi)$
be the empirical risk and consider the Lasso estimator
\begin{equation}
  \hat{\bgamma} \in \arg\min_{\bgamma}
    \left\{ \mathcal{L}_n(\bgamma) + \lambda_n \Psi_p(\bgamma) \right\},
  \qquad \lambda_n = \lambda/n,
  \label{eq:avg_scale}
\end{equation}
with $\Psi_p(\bgamma) = \|\bgamma_{-0}\|_1$, which is
\eqref{eq:penloglik_modal} divided by $-n$. The Hessian of $\mathcal{L}_n$ is
$n^{-1}\X^\top \W(\bgamma) \X$, with
$\W(\bgamma) = \mathrm{diag}(w_1, \ldots, w_n)$ the working weights of
\eqref{eq:scorecurv}, which are positive whenever $y_i > 0$, so that
$\mathcal{L}_n$ is convex. Since the convexity is not strict when $p + 1 > n$,
we assume, as in \cite{negahban2012}, restricted strong convexity (RSC)
locally over a cone of approximately sparse directions. Let $\mathcal{S}$ be
the set of active covariates, $s = |\mathcal{S}|$,
$\bar{\mathcal{S}} = \mathcal{S} \cup \{0\}$ and
$\mathcal{C} = \{\Delta \in \mathbb{R}^{p+1} :
  \|\Delta_{\bar{\mathcal{S}}^c}\|_1 \le 3\|\Delta_{\bar{\mathcal{S}}}\|_1\}$.

\begin{itemize}
  \item[(A1)] \textit{(RSC)} There exist $\kappa > 0$ and $r > 0$ such that
    $\mathcal{L}_n(\bgamma^0 + \Delta) - \mathcal{L}_n(\bgamma^0)
      - \nabla\mathcal{L}_n(\bgamma^0)^\top \Delta
      \ge \tfrac{\kappa}{2}\|\Delta\|_2^2$
    for all $\Delta \in \mathcal{C}$ with $\|\Delta\|_2 \le r$.
  \item[(A2)] \textit{(Score domination)}
    $\lambda_n \ge 2\|\nabla\mathcal{L}_n(\bgamma^0)\|_\infty$.
\end{itemize}

\begin{theorem}[Estimation error of the modal Lasso]
\label{thm:consistency}
Let $\hat{\bgamma}$ be a minimizer of \eqref{eq:avg_scale} and suppose
that (A1) and (A2) hold with $3\sqrt{s+1}\,\lambda_n/\kappa < r$. Then
\begin{equation}
    \|\hat{\bgamma} - \bgamma^0\|_2 \;\le\;
      \frac{3\sqrt{s+1}}{\kappa}\,\lambda_n .
\end{equation}
Consequently, if $\lambda_n \asymp \sqrt{\log(p)/n}$, then
$\|\hat{\bgamma} - \bgamma^0\|_2 = \mathcal{O}_P\!\left(\sqrt{s\log(p)/n}\right)$,
which tends to zero whenever $s\log(p)/n \to 0$.
\end{theorem}

\begin{proof}
Let $\hat\Delta = \hat{\bgamma} - \bgamma^0$ and
$F(\Delta) = \mathcal{L}_n(\bgamma^0 + \Delta) - \mathcal{L}_n(\bgamma^0)
  + \lambda_n\{\Psi_p(\bgamma^0 + \Delta) - \Psi_p(\bgamma^0)\}$, which is
convex and satisfies $F(\bm{0}) = 0$ and $F(\hat\Delta) \le 0$. Since the
slopes of $\bgamma^0$ are supported on $\mathcal{S}$, the triangle inequality
gives
$\Psi_p(\bgamma^0 + \Delta) - \Psi_p(\bgamma^0)
  \ge \|\Delta_{\bar{\mathcal{S}}^c}\|_1 - \|\Delta_{\mathcal{S}}\|_1$,
and by H\"older's inequality and (A2),
$|\nabla\mathcal{L}_n(\bgamma^0)^\top\Delta| \le \tfrac{\lambda_n}{2}\|\Delta\|_1$.
By convexity of $\mathcal{L}_n$,
$\mathcal{L}_n(\bgamma^0 + \Delta) - \mathcal{L}_n(\bgamma^0)
  \ge \nabla\mathcal{L}_n(\bgamma^0)^\top\Delta$, so that
\[
  0 \ge F(\hat\Delta) \ge \frac{\lambda_n}{2}
    \left(\|\hat\Delta_{\bar{\mathcal{S}}^c}\|_1
      - 3\|\hat\Delta_{\bar{\mathcal{S}}}\|_1\right),
\]
that is, $\hat\Delta \in \mathcal{C}$. On the other hand, for every
$\Delta \in \mathcal{C}$ with $\|\Delta\|_2 = \delta \le r$, (A1) and the same
two bounds yield
\begin{equation}
  F(\Delta) \ge \frac{\kappa}{2}\delta^2
    - \frac{3\lambda_n}{2}\|\Delta_{\bar{\mathcal{S}}}\|_1
    + \frac{\lambda_n}{2}\|\Delta_{\bar{\mathcal{S}}^c}\|_1
  \ge \frac{\kappa}{2}\delta^2 - \frac{3\lambda_n\sqrt{s+1}}{2}\,\delta ,
  \label{eq:cone_ineq}
\end{equation}
which is strictly positive whenever
$\delta > \delta^\ast = 3\sqrt{s+1}\,\lambda_n/\kappa$. Suppose that
$\|\hat\Delta\|_2 > \delta$ for some $\delta \in (\delta^\ast, r]$. Then
$t\hat\Delta$, with $t = \delta/\|\hat\Delta\|_2 < 1$, belongs to
$\mathcal{C}$ and has norm $\delta$, whereas the convexity of $F$ gives
$F(t\hat\Delta) \le tF(\hat\Delta) + (1-t)F(\bm{0}) \le 0$, which contradicts
\eqref{eq:cone_ineq}. Hence $\|\hat\Delta\|_2 \le \delta$ for every
$\delta \in (\delta^\ast, r]$, and the claim follows.
\end{proof}

\begin{remark}[Condition (A2)]
\label{rem:tails}
Since $\nabla\mathcal{L}_n(\bgamma^0) = -n^{-1}\sum_i s_i \bx_i$, where
the scores $s_i$ of \eqref{eq:scorecurv} are Gaussian for the Lognormal family
and centred Gamma and exponential variates for the Gamma and Weibull families,
for bounded covariates (A2) holds with $\lambda_n \asymp \sqrt{\log(p)/n}$ with
high probability, by Bernstein's inequality and a union bound, provided that
$\log p = o(n)$.
\end{remark}

\begin{remark}[Scope of the result]
\label{rem:scope}
Since the objective in \eqref{eq:avg_scale} is convex,
Theorem~\ref{thm:consistency} applies to the limit of
Algorithm~\ref{alg:emlasso} \cite{lee2014}. The bound contains no smoothing
term because \eqref{eq:mstep} is solved exactly, whereas replacing
$|\gamma_j|$ by a differentiable surrogate adds a bias that grows with $p$.
The joint analysis with an estimated $\phi$ could follow \cite{loh2015} and is
not pursued here.
\end{remark}

\section{Simulation Studies}
\label{sec:simulation}

We evaluate the finite-sample performance of the proposed method in a
main design with independent covariates, in which every active coefficient
equals one (Sections~\ref{sec:design}--\ref{sec:gamma_study}), and in a design
with correlated and binary covariates (Section~\ref{sec:structure_study}).
Each configuration is replicated $B = 500$ times.

\subsection{Design}
\label{sec:design}

For each replication, covariates are generated independently as
$x_{ij} \stackrel{iid}{\sim} \mathrm{U}(0,1)$ for $j = 1, \ldots, p$, and the
design matrix $\X$ includes a leading column of ones for the intercept. The
true coefficient vector has the sparse form
$\bgamma^0 = (\gamma_0^0, \gamma_1^0, \ldots, \gamma_{p^*}^0, 0, \ldots, 0)^\top$,
where $p^*$ is the number of active covariates, with $\gamma_0^0 = 0.8$
and $\gamma_j^0 = 1$ for $j = 1, \ldots, p^*$. The conditional mode is
$M_i = \exp(\bx_i^\top\bgamma^0)$ and responses are drawn from the
family-specific distribution with true dispersion $\phi^0 = 0.3$. All results
use $\gamma_{\mathrm{EBIC}} = 0.5$, except in Section~\ref{sec:gamma_study}.
Reproducibility follows a fixed scheme: a master seed of $2025$ with
replication $b$ seeded at $2025 + b$, so that all methods and all scenarios
are compared on identical data. All fits use Algorithm~\ref{alg:emlasso}
with the line search, as implemented in the \texttt{R} package
\texttt{ModalNet} \cite{modalnet}, version 0.3.1, and the paths of every method are stopped
once more than $\lfloor n/\log n \rfloor$ covariates are active, a limit that
exceeds $p^*$ in every scenario.

Five scenarios are considered, which differ in the dimension and in the
number of active covariates: S1 with $p = 20$ and $p^* = 4$, and S2 with
$p = 50$ and $p^* = 5$, both with $n \in \{100, 200, 400\}$; S3 and S4 with
$p = 100$ and $p = 500$, respectively, and $p^* = 5$; and S5 with $p = 1000$
and $p^* = 10$, the last three with $n \in \{50, 100, 200\}$, so that
$p \ge n$ throughout.

Following \cite{chun2010} and \cite{oliveira2025}, we evaluate performance
through two groups of metrics.

\begin{itemize}
  \item \textbf{Variable selection metrics:} sensitivity, specificity and
    accuracy. Sensitivity is the proportion of truly nonzero coefficients
    among the first $p^*$ components that are estimated as nonzero, with
    no threshold, since the update \eqref{eq:mstep} returns exact zeros;
    specificity is the proportion of truly zero
    coefficients among the remaining $p - p^*$ that are estimated as zero; and
    accuracy is the overall proportion of correctly classified coefficients
    out of $p$. The intercept is excluded from all three. We also report the
    average model size $|\hat{\mathcal{S}}|$.
  \item \textbf{Estimation error:} the root mean squared error, averaged over
    replications,
    \begin{equation}
      \mathrm{RMSE} = \frac{1}{B}\sum_{b=1}^{B}
        \sqrt{\frac{1}{p+1}\sum_{j=0}^{p}
          \left(\hat\gamma_j^{(b)} - \gamma_j^0\right)^2},
    \end{equation}
    where the inner sum runs over all $p+1$ coefficients, including the
    intercept.
\end{itemize}

As a benchmark we include the \emph{log-Gaussian Lasso}: an ordinary Gaussian
Lasso fit to $\log \y$, tuned by the same EBIC criterion. To make the criteria
comparable across scales, its log-likelihood is mapped back to the scale of
$y$ through the change-of-variables Jacobian, $|dy/d(\log y)| = 1/y$, so that
$\ell_Y = \ell_{\log Y} - \sum_i \log y_i$. This competitor is the natural
practical alternative to modal regression, and by
Proposition~\ref{prop:lognormal} it is also a sharp diagnostic: the two
procedures must agree for Lognormal responses and are free to differ
elsewhere.

\subsection{Results}
\label{sec:results}

Table~\ref{tab:sim_selection} reports sensitivity, specificity and accuracy;
Table~\ref{tab:sim_rmse} reports RMSE and average model size. Every cell is
based on $B = 500$ replications, and the selected fit converged in all
but one of the $22{,}500$ replications. In the largest configuration
($p = 1000$, $n = 200$), a complete path with EBIC selection took on average
$1.1$~s for the Gamma family, $1.3$~s for the Weibull and $1.0$~s for the
Lognormal, on one core of an AMD EPYC 9554 processor.

\begin{table}[!ht]
  \caption{Variable selection performance ($B = 500$, $\phi^0 = 0.3$, $\gamma_{\mathrm{EBIC}} = 0.5$, active coefficients equal to $1$).}
  \label{tab:sim_selection}
  \centering
  \scriptsize
  \setlength{\tabcolsep}{3.5pt}
  \begin{tabular}{ll l ccc ccc ccc}
\toprule
& & & \multicolumn{3}{c}{\textbf{Gamma}} & \multicolumn{3}{c}{\textbf{Weibull}} & \multicolumn{3}{c}{\textbf{Lognormal}} \\
\cmidrule(lr){4-6} \cmidrule(lr){7-9} \cmidrule(lr){10-12}
\textbf{Scenario} & $n$ & Method & Sens & Spec & Acc & Sens & Spec & Acc & Sens & Spec & Acc \\
\midrule
\multirow{6}{*}{\shortstack[l]{S1: Low-dim.\\($p=20$, $p^*=4$)}} & \multirow{2}{*}{100} & Modal Lasso & 0.995 & 0.943 & 0.953 & 0.526 & 0.979 & 0.888 & 0.968 & 0.953 & 0.956 \\
 &  & Log-Gaussian Lasso & 0.984 & 0.944 & 0.952 & 0.244 & 0.992 & 0.843 & 0.968 & 0.953 & 0.956 \\
\addlinespace[2pt]
 & \multirow{2}{*}{200} & Modal Lasso & 1.000 & 0.953 & 0.962 & 0.985 & 0.950 & 0.957 & 1.000 & 0.959 & 0.967 \\
 &  & Log-Gaussian Lasso & 1.000 & 0.957 & 0.965 & 0.801 & 0.966 & 0.933 & 1.000 & 0.959 & 0.967 \\
\addlinespace[2pt]
 & \multirow{2}{*}{400} & Modal Lasso & 1.000 & 0.964 & 0.971 & 1.000 & 0.958 & 0.967 & 1.000 & 0.961 & 0.968 \\
 &  & Log-Gaussian Lasso & 1.000 & 0.969 & 0.975 & 0.998 & 0.961 & 0.969 & 1.000 & 0.961 & 0.968 \\
\cmidrule{1-12}
\multirow{6}{*}{\shortstack[l]{S2: Mod-dim.\\($p=50$, $p^*=5$)}} & \multirow{2}{*}{100} & Modal Lasso & 0.952 & 0.971 & 0.970 & 0.242 & 0.996 & 0.921 & 0.851 & 0.981 & 0.968 \\
 &  & Log-Gaussian Lasso & 0.934 & 0.976 & 0.972 & 0.095 & 0.999 & 0.909 & 0.851 & 0.981 & 0.968 \\
\addlinespace[2pt]
 & \multirow{2}{*}{200} & Modal Lasso & 1.000 & 0.977 & 0.979 & 0.915 & 0.978 & 0.972 & 1.000 & 0.977 & 0.979 \\
 &  & Log-Gaussian Lasso & 1.000 & 0.979 & 0.981 & 0.587 & 0.991 & 0.951 & 1.000 & 0.977 & 0.979 \\
\addlinespace[2pt]
 & \multirow{2}{*}{400} & Modal Lasso & 1.000 & 0.981 & 0.983 & 1.000 & 0.979 & 0.981 & 1.000 & 0.983 & 0.985 \\
 &  & Log-Gaussian Lasso & 1.000 & 0.983 & 0.984 & 0.996 & 0.983 & 0.985 & 1.000 & 0.983 & 0.985 \\
\cmidrule{1-12}
\multirow{6}{*}{\shortstack[l]{S3: High-dim.\\($p=100$, $p^*=5$)}} & \multirow{2}{*}{50} & Modal Lasso & 0.118 & 0.999 & 0.955 & 0.023 & 1.000 & 0.951 & 0.061 & 1.000 & 0.953 \\
 &  & Log-Gaussian Lasso & 0.081 & 1.000 & 0.954 & 0.010 & 1.000 & 0.950 & 0.061 & 1.000 & 0.953 \\
\addlinespace[2pt]
 & \multirow{2}{*}{100} & Modal Lasso & 0.840 & 0.988 & 0.981 & 0.141 & 0.999 & 0.956 & 0.624 & 0.995 & 0.976 \\
 &  & Log-Gaussian Lasso & 0.800 & 0.991 & 0.982 & 0.043 & 1.000 & 0.952 & 0.624 & 0.995 & 0.976 \\
\addlinespace[2pt]
 & \multirow{2}{*}{200} & Modal Lasso & 1.000 & 0.990 & 0.991 & 0.788 & 0.992 & 0.982 & 1.000 & 0.989 & 0.990 \\
 &  & Log-Gaussian Lasso & 1.000 & 0.991 & 0.991 & 0.347 & 0.998 & 0.966 & 1.000 & 0.989 & 0.990 \\
\cmidrule{1-12}
\multirow{6}{*}{\shortstack[l]{S4: Very high\\($p=500$, $p^*=5$)}} & \multirow{2}{*}{50} & Modal Lasso & 0.030 & 1.000 & 0.990 & 0.005 & 1.000 & 0.990 & 0.009 & 1.000 & 0.990 \\
 &  & Log-Gaussian Lasso & 0.020 & 1.000 & 0.990 & 0.002 & 1.000 & 0.990 & 0.009 & 1.000 & 0.990 \\
\addlinespace[2pt]
 & \multirow{2}{*}{100} & Modal Lasso & 0.368 & 0.999 & 0.993 & 0.034 & 1.000 & 0.990 & 0.179 & 1.000 & 0.992 \\
 &  & Log-Gaussian Lasso & 0.300 & 1.000 & 0.993 & 0.013 & 1.000 & 0.990 & 0.179 & 1.000 & 0.992 \\
\addlinespace[2pt]
 & \multirow{2}{*}{200} & Modal Lasso & 1.000 & 0.998 & 0.998 & 0.394 & 1.000 & 0.993 & 0.990 & 0.998 & 0.998 \\
 &  & Log-Gaussian Lasso & 1.000 & 0.998 & 0.998 & 0.095 & 1.000 & 0.991 & 0.990 & 0.998 & 0.998 \\
\cmidrule{1-12}
\multirow{6}{*}{\shortstack[l]{S5: Extreme\\($p=1000$, $p^*=10$)}} & \multirow{2}{*}{50} & Modal Lasso & 0.003 & 1.000 & 0.990 & 0.002 & 1.000 & 0.990 & 0.001 & 1.000 & 0.990 \\
 &  & Log-Gaussian Lasso & 0.001 & 1.000 & 0.990 & 0.000 & 1.000 & 0.990 & 0.001 & 1.000 & 0.990 \\
\addlinespace[2pt]
 & \multirow{2}{*}{100} & Modal Lasso & 0.017 & 1.000 & 0.990 & 0.018 & 1.000 & 0.990 & 0.009 & 1.000 & 0.990 \\
 &  & Log-Gaussian Lasso & 0.007 & 1.000 & 0.990 & 0.001 & 1.000 & 0.990 & 0.009 & 1.000 & 0.990 \\
\addlinespace[2pt]
 & \multirow{2}{*}{200} & Modal Lasso & 0.864 & 0.997 & 0.996 & 0.127 & 1.000 & 0.991 & 0.695 & 0.999 & 0.996 \\
 &  & Log-Gaussian Lasso & 0.836 & 0.998 & 0.997 & 0.015 & 1.000 & 0.990 & 0.695 & 0.999 & 0.996 \\
\bottomrule
\end{tabular}
\end{table}

\begin{table}[!ht]
  \caption{Estimation error and model size ($B = 500$). The true model sizes are $p^* = 4$ (S1), $p^* = 5$ (S2--S4) and $p^* = 10$ (S5). Since the RMSE is normalized by $p+1$, it is not comparable across scenarios; see Table~\ref{tab:l2}.}
  \label{tab:sim_rmse}
  \centering
  \scriptsize
  \setlength{\tabcolsep}{4pt}
  \begin{tabular}{ll l cc cc cc}
\toprule
& & & \multicolumn{2}{c}{\textbf{Gamma}} & \multicolumn{2}{c}{\textbf{Weibull}} & \multicolumn{2}{c}{\textbf{Lognormal}} \\
\cmidrule(lr){4-5} \cmidrule(lr){6-7} \cmidrule(lr){8-9}
\textbf{Scenario} & $n$ & Method & RMSE & $|\hat{\mathcal{S}}|$ & RMSE & $|\hat{\mathcal{S}}|$ & RMSE & $|\hat{\mathcal{S}}|$ \\
\midrule
\multirow{6}{*}{\shortstack[l]{S1: Low-dim.\\($p=20$, $p^*=4$)}} & \multirow{2}{*}{100} & Modal Lasso & 0.203 & 4.9 & 0.378 & 2.4 & 0.235 & 4.6 \\
 &  & Log-Gaussian Lasso & 0.238 & 4.8 & 0.632 & 1.1 & 0.285 & 4.6 \\
\addlinespace[2pt]
 & \multirow{2}{*}{200} & Modal Lasso & 0.139 & 4.8 & 0.224 & 4.7 & 0.157 & 4.7 \\
 &  & Log-Gaussian Lasso & 0.169 & 4.7 & 0.438 & 3.8 & 0.207 & 4.7 \\
\addlinespace[2pt]
 & \multirow{2}{*}{400} & Modal Lasso & 0.098 & 4.6 & 0.153 & 4.7 & 0.112 & 4.6 \\
 &  & Log-Gaussian Lasso & 0.127 & 4.5 & 0.309 & 4.6 & 0.162 & 4.6 \\
\cmidrule{1-9}
\multirow{6}{*}{\shortstack[l]{S2: Mod-dim.\\($p=50$, $p^*=5$)}} & \multirow{2}{*}{100} & Modal Lasso & 0.197 & 6.0 & 0.319 & 1.4 & 0.234 & 5.1 \\
 &  & Log-Gaussian Lasso & 0.225 & 5.8 & 0.516 & 0.5 & 0.274 & 5.1 \\
\addlinespace[2pt]
 & \multirow{2}{*}{200} & Modal Lasso & 0.131 & 6.0 & 0.203 & 5.6 & 0.146 & 6.0 \\
 &  & Log-Gaussian Lasso & 0.155 & 5.9 & 0.395 & 3.3 & 0.180 & 6.0 \\
\addlinespace[2pt]
 & \multirow{2}{*}{400} & Modal Lasso & 0.090 & 5.9 & 0.140 & 5.9 & 0.101 & 5.8 \\
 &  & Log-Gaussian Lasso & 0.111 & 5.8 & 0.259 & 5.7 & 0.135 & 5.8 \\
\cmidrule{1-9}
\multirow{6}{*}{\shortstack[l]{S3: High-dim.\\($p=100$, $p^*=5$)}} & \multirow{2}{*}{50} & Modal Lasso & 0.289 & 0.7 & 0.251 & 0.1 & 0.296 & 0.3 \\
 &  & Log-Gaussian Lasso & 0.330 & 0.4 & 0.382 & 0.1 & 0.345 & 0.3 \\
\addlinespace[2pt]
 & \multirow{2}{*}{100} & Modal Lasso & 0.175 & 5.3 & 0.234 & 0.8 & 0.213 & 3.6 \\
 &  & Log-Gaussian Lasso & 0.202 & 4.8 & 0.376 & 0.2 & 0.248 & 3.6 \\
\addlinespace[2pt]
 & \multirow{2}{*}{200} & Modal Lasso & 0.107 & 5.9 & 0.167 & 4.7 & 0.117 & 6.0 \\
 &  & Log-Gaussian Lasso & 0.125 & 5.9 & 0.329 & 1.9 & 0.143 & 6.0 \\
\cmidrule{1-9}
\multirow{6}{*}{\shortstack[l]{S4: Very high\\($p=500$, $p^*=5$)}} & \multirow{2}{*}{50} & Modal Lasso & 0.134 & 0.2 & 0.113 & 0.0 & 0.136 & 0.0 \\
 &  & Log-Gaussian Lasso & 0.153 & 0.1 & 0.173 & 0.0 & 0.159 & 0.0 \\
\addlinespace[2pt]
 & \multirow{2}{*}{100} & Modal Lasso & 0.116 & 2.1 & 0.108 & 0.2 & 0.127 & 1.0 \\
 &  & Log-Gaussian Lasso & 0.134 & 1.7 & 0.172 & 0.1 & 0.148 & 1.0 \\
\addlinespace[2pt]
 & \multirow{2}{*}{200} & Modal Lasso & 0.060 & 6.0 & 0.094 & 2.2 & 0.067 & 5.7 \\
 &  & Log-Gaussian Lasso & 0.069 & 5.9 & 0.167 & 0.5 & 0.080 & 5.7 \\
\cmidrule{1-9}
\multirow{6}{*}{\shortstack[l]{S5: Extreme\\($p=1000$, $p^*=10$)}} & \multirow{2}{*}{50} & Modal Lasso & 0.152 & 0.0 & 0.118 & 0.1 & 0.166 & 0.0 \\
 &  & Log-Gaussian Lasso & 0.191 & 0.0 & 0.205 & 0.0 & 0.195 & 0.0 \\
\addlinespace[2pt]
 & \multirow{2}{*}{100} & Modal Lasso & 0.147 & 0.2 & 0.115 & 0.2 & 0.165 & 0.1 \\
 &  & Log-Gaussian Lasso & 0.190 & 0.1 & 0.205 & 0.0 & 0.194 & 0.1 \\
\addlinespace[2pt]
 & \multirow{2}{*}{200} & Modal Lasso & 0.095 & 11.7 & 0.109 & 1.4 & 0.114 & 8.3 \\
 &  & Log-Gaussian Lasso & 0.109 & 10.2 & 0.203 & 0.2 & 0.131 & 8.3 \\
\bottomrule
\end{tabular}
\end{table}

\begin{table}[!ht]
  \caption{Unscaled estimation error $\|\hat{\bgamma} - \bgamma^0\|_2$ at
  the largest $n$ in each scenario, for the modal Lasso and the log-Gaussian
  Lasso. Unlike the RMSE, this quantity is comparable across $p$.}
  \label{tab:l2}
  \centering
  \footnotesize
  \begin{tabular}{l cc cc cc}
\toprule
& \multicolumn{2}{c}{\textbf{Gamma}} & \multicolumn{2}{c}{\textbf{Weibull}} & \multicolumn{2}{c}{\textbf{Lognormal}} \\
\cmidrule(lr){2-3} \cmidrule(lr){4-5} \cmidrule(lr){6-7}
Scenario & Modal & Log-G. & Modal & Log-G. & Modal & Log-G. \\
\midrule
S1 ($p=20$, $n=400$) & 0.450 & 0.582 & 0.701 & 1.415 & 0.512 & 0.744 \\
S2 ($p=50$, $n=400$) & 0.645 & 0.792 & 1.000 & 1.852 & 0.718 & 0.963 \\
S3 ($p=100$, $n=200$) & 1.078 & 1.252 & 1.680 & 3.302 & 1.179 & 1.435 \\
S4 ($p=500$, $n=200$) & 1.354 & 1.546 & 2.112 & 3.729 & 1.510 & 1.793 \\
S5 ($p=1000$, $n=200$) & 3.005 & 3.463 & 3.452 & 6.437 & 3.606 & 4.148 \\
\bottomrule
\end{tabular}
\end{table}

The modal Lasso detects the active covariates once the sample size is
moderate relative to the dimension: its average sensitivity is $0.451$ at
$n = 100$, $0.850$ at $n = 200$ and $1.000$ at $n = 400$, and it is at least
$0.9$ in $19$ of the $45$ cells. For Gamma responses with $n = 200$, for
instance, the sensitivity equals $1.000$ with $p = 100$ and $p = 500$, and
$0.864$ with $p = 1000$. The sample size is the limiting factor at $n = 50$,
where the average sensitivity is $0.028$ and both methods select less than
one covariate on average. The specificity lies between $0.943$ and $1.000$; it is
lowest in S1 and S2, where the term in $\log p$ of the EBIC \eqref{eq:ebic} is
smallest.

For Gamma and Weibull responses, the sensitivity of the modal Lasso is at
least that of the log-Gaussian Lasso in the $30$ cells and strictly higher in
$25$, the five ties being cells in which both methods attain $1.000$, and its
RMSE is lower in the $30$ cells. The largest gains occur for the Weibull
family, for which the log-Gaussian working model is most misspecified: the
sensitivities are $0.915$ and $0.587$ in S2, $0.788$ and $0.347$ in S3, and
$0.394$ and $0.095$ in S4, all with $n = 200$, and the RMSEs $0.378$ and
$0.632$ in S1 with $n = 100$. The modal fit selects slightly more covariates,
so that its specificity is at most $0.016$ below that of the benchmark, as in
S1 with Weibull responses and $n = 200$, where they equal $0.950$ and $0.966$.
For Lognormal responses, the two procedures select models of the same size in
the $7{,}500$ replications, so that sensitivity, specificity and accuracy
coincide in the fifteen cells, in agreement with
Proposition~\ref{prop:lognormal}. Their RMSEs differ, for instance $0.112$
against $0.162$ in S1 with $n = 400$, because the intercept of the
log-Gaussian Lasso estimates $\gamma_0^0 + \phi^0$ instead of $\gamma_0^0$.

The RMSE divides the squared error by $p+1$, so that it is not comparable
across scenarios. Table~\ref{tab:l2} reports the unscaled error, which
increases with $p$, for instance from $0.450$ in S1 to $3.005$ in S5 for the
Gamma family, and is lower for the modal Lasso than for the benchmark in the
fifteen cells, with the largest differences for the Weibull family, for
instance $1.680$ against $3.302$ in S3.

For the Weibull family, the dispersion of the selected fit lies at its lower
limit $\phi = 10^{-4}$, that is, at the exponential limit, in $37.7\%$ of the
replications. This occurs only when active covariates are missed: in $64.7\%$
of the replications in which the selected model is empty and in none of the
$2{,}627$ in which every active covariate is selected, where the mean of
$\hat\phi$ is $0.274$. The part of the linear predictor that is not modelled
makes the response more dispersed than a Weibull density with shape
$\phi^0 + 1.01 = 1.31$, so that the fit moves towards the exponential limit.
The Gamma fit reaches its exponential limit, $\phi = 10^{4}$, in $2.9\%$ of the
replications, all of them in S5 with $n \le 100$.

With weaker active coefficients, of magnitude between $0.2$ and $0.6$,
both methods need larger samples, but the comparison between them is
unchanged: for Gamma and Weibull responses the modal Lasso attains at least the
sensitivity of the log-Gaussian Lasso in the $30$ cells and a lower RMSE in
$28$, with specificity above $0.98$. On Scenarios~S2 and~S3 of this design
($n \in \{100, 200\}$, $B = 200$), fitting a family other than the generating
one reduced the sensitivity by at most $0.042$, the adaptive Lasso, with
weights from an initial modal Lasso fit, raised it from $0.107$ to $0.283$,
and the elastic net with $\alpha = 0.5$ gave $0.087$, as expected with
independent covariates.

\subsection{Sensitivity to the EBIC parameter}
\label{sec:gamma_study}

The results above use $\gamma_{\mathrm{EBIC}} = 0.5$. Since
$\gamma_{\mathrm{EBIC}}$ enters only through the selection of $\lambda$ from
an already computed path, the fits selected under
$\gamma_{\mathrm{EBIC}} \in \{0, 0.25, 0.5, 1\}$ are read from the paths of the
main design, so that Table~\ref{tab:gamma_sweep} is based on the same $500$
replications. It reports Scenarios~S3--S5 with $n \in \{100, 200\}$.

\begin{table}[!ht]
  \caption{Sensitivity and specificity as a function of $\gamma_{\mathrm{EBIC}}$
  in the main design ($B = 500$), read from the same paths.}
  \label{tab:gamma_sweep}
  \centering
  \footnotesize
  \setlength{\tabcolsep}{4.5pt}
  \begin{tabular}{lll cc cc cc cc}
\toprule
& & & \multicolumn{2}{c}{$\gamma_{\mathrm{EBIC}} = 0$} & \multicolumn{2}{c}{$0.25$} & \multicolumn{2}{c}{$0.5$} & \multicolumn{2}{c}{$1$} \\
\cmidrule(lr){4-5} \cmidrule(lr){6-7} \cmidrule(lr){8-9} \cmidrule(lr){10-11}
Scenario & $n$ & Family & Sens & Spec & Sens & Spec & Sens & Spec & Sens & Spec \\
\midrule
\multirow{6}{*}{\shortstack[l]{S3\\($p=100$)}} & \multirow{3}{*}{100} & Gamma & 0.991 & 0.957 & 0.954 & 0.978 & 0.840 & 0.988 & 0.442 & 0.998 \\
 &  & Weibull & 0.580 & 0.979 & 0.295 & 0.996 & 0.141 & 0.999 & 0.025 & 1.000 \\
 &  & Lognormal & 0.974 & 0.964 & 0.854 & 0.985 & 0.624 & 0.995 & 0.146 & 1.000 \\
\addlinespace[2pt]
 & \multirow{3}{*}{200} & Gamma & 1.000 & 0.971 & 1.000 & 0.986 & 1.000 & 0.990 & 0.998 & 0.994 \\
 &  & Weibull & 0.980 & 0.969 & 0.914 & 0.985 & 0.788 & 0.992 & 0.372 & 0.999 \\
 &  & Lognormal & 1.000 & 0.974 & 1.000 & 0.986 & 1.000 & 0.989 & 0.995 & 0.994 \\
\cmidrule{1-11}
\multirow{6}{*}{\shortstack[l]{S4\\($p=500$)}} & \multirow{3}{*}{100} & Gamma & 0.916 & 0.990 & 0.647 & 0.998 & 0.368 & 0.999 & 0.052 & 1.000 \\
 &  & Weibull & 0.268 & 0.998 & 0.102 & 1.000 & 0.034 & 1.000 & 0.004 & 1.000 \\
 &  & Lognormal & 0.822 & 0.993 & 0.500 & 0.999 & 0.179 & 1.000 & 0.007 & 1.000 \\
\addlinespace[2pt]
 & \multirow{3}{*}{200} & Gamma & 1.000 & 0.993 & 1.000 & 0.997 & 1.000 & 0.998 & 0.973 & 0.999 \\
 &  & Weibull & 0.870 & 0.994 & 0.661 & 0.998 & 0.394 & 1.000 & 0.069 & 1.000 \\
 &  & Lognormal & 0.999 & 0.994 & 0.998 & 0.997 & 0.990 & 0.998 & 0.864 & 0.999 \\
\cmidrule{1-11}
\multirow{6}{*}{\shortstack[l]{S5\\($p=1000$)}} & \multirow{3}{*}{100} & Gamma & 0.416 & 0.996 & 0.081 & 1.000 & 0.017 & 1.000 & 0.002 & 1.000 \\
 &  & Weibull & 0.147 & 0.998 & 0.043 & 1.000 & 0.018 & 1.000 & 0.003 & 1.000 \\
 &  & Lognormal & 0.297 & 0.998 & 0.038 & 1.000 & 0.009 & 1.000 & 0.001 & 1.000 \\
\addlinespace[2pt]
 & \multirow{3}{*}{200} & Gamma & 0.999 & 0.989 & 0.982 & 0.995 & 0.864 & 0.997 & 0.258 & 1.000 \\
 &  & Weibull & 0.668 & 0.994 & 0.308 & 0.999 & 0.127 & 1.000 & 0.022 & 1.000 \\
 &  & Lognormal & 0.998 & 0.992 & 0.947 & 0.996 & 0.695 & 0.999 & 0.056 & 1.000 \\
\bottomrule
\end{tabular}
\end{table}

Lowering $\gamma_{\mathrm{EBIC}}$ from $0.5$ to $0$ increases the
sensitivity in every cell in which it is below one, at a cost in specificity
of at most $0.031$, so that the specificity remains at least $0.957$. The gain
is largest when $n$ is small relative to $p$: for Gamma responses the
sensitivity rises from $0.368$ to $0.916$ in S4 with $n = 100$ and from
$0.864$ to $0.999$ in S5 with $n = 200$, and for Weibull responses from
$0.394$ to $0.870$ in S4 and from $0.127$ to $0.668$ in S5, both with
$n = 200$. Hence, when the purpose is screening with $p$ of the order of
hundreds or thousands, $\gamma_{\mathrm{EBIC}} = 0$, that is, the BIC over
paths restricted to at most $\lfloor n/\log n \rfloor$ active covariates, is
preferable to $\gamma_{\mathrm{EBIC}} = 0.5$, which should be kept when false
positives are costly.

\subsection{Correlated and binary covariates}
\label{sec:structure_study}

We also consider a structured design with $p \in \{50, 100, 500\}$ and
$p^* = 5$, in which $40\%$ of the covariates are drawn from a multivariate
normal distribution with $\mathrm{Cor}(x_{ij}, x_{ik}) = \rho^{|j-k|}$, $20\%$
are independent $\mathrm{Bernoulli}(0.5)$ and the remaining $40\%$ are
independent standard normal. Two active covariates are placed in the
correlated block, one in the binary block and two in the independent block,
with $\rho \in \{0, 0.5, 0.8\}$ and $n \in \{50, 100, 200\}$; the five
active coefficients are equal to $1$ and $\gamma_0^0 = 0.8$. Since the
covariates are no longer $\mathrm{U}(0,1)$, the standard deviation of the
linear predictor ranges from $2.04$ to $2.42$, against $0.58$ to
$0.91$ in the main design, so that the sensitivities in
Table~\ref{tab:structure} are not comparable with those of
Table~\ref{tab:sim_selection}; the purpose here is to isolate the effect of
the correlation and of the type of covariate.

\begin{table}[!ht]
  \caption{Selection performance under correlated and binary
    covariates, averaged over the three response families ($B = 500$ per cell,
    $\phi^0 = 0.3$, $\gamma_{\mathrm{EBIC}} = 0.5$). Block sizes are given as
    (correlated/binary/independent). The correlation $\rho$ applies within the
    correlated block only.}
  \label{tab:structure}
  \centering
  \scriptsize
  \setlength{\tabcolsep}{4pt}
  \begin{tabular}{ll cc cc cc}
\toprule
& & \multicolumn{2}{c}{$\rho = 0$} & \multicolumn{2}{c}{$\rho = 0.5$} & \multicolumn{2}{c}{$\rho = 0.8$} \\
\cmidrule(lr){3-4} \cmidrule(lr){5-6} \cmidrule(lr){7-8}
\textbf{Configuration} & $n$ & Sens & Spec & Sens & Spec & Sens & Spec \\
\midrule
\multirow{3}{*}{\shortstack[l]{C1: $p=50$\\(20/10/20)}} & 50 & 0.972 & 0.951 & 0.979 & 0.960 & 0.980 & 0.959 \\
 & 100 & 0.999 & 0.967 & 0.999 & 0.972 & 0.999 & 0.969 \\
 & 200 & 1.000 & 0.979 & 1.000 & 0.980 & 1.000 & 0.976 \\
\cmidrule{1-8}
\multirow{3}{*}{\shortstack[l]{C2: $p=100$\\(40/20/40)}} & 50 & 0.929 & 0.974 & 0.954 & 0.978 & 0.955 & 0.978 \\
 & 100 & 0.998 & 0.984 & 0.998 & 0.987 & 0.998 & 0.986 \\
 & 200 & 1.000 & 0.990 & 1.000 & 0.991 & 1.000 & 0.989 \\
\cmidrule{1-8}
\multirow{3}{*}{\shortstack[l]{C3: $p=500$\\(200/100/200)}} & 50 & 0.593 & 0.995 & 0.781 & 0.995 & 0.807 & 0.996 \\
 & 100 & 0.988 & 0.996 & 0.990 & 0.997 & 0.992 & 0.997 \\
 & 200 & 1.000 & 0.998 & 1.000 & 0.998 & 1.000 & 0.998 \\
\bottomrule
\end{tabular}
\end{table}

Correlation within the block does not degrade selection: averaged over
families, dimensions and sample sizes, the sensitivity is $0.942$ at
$\rho = 0$, $0.967$ at $\rho = 0.5$ and $0.970$ at $\rho = 0.8$, with the
largest difference at C3 with $n = 50$ ($0.593$ against $0.807$), and the
specificity is unaffected. This holds for an autoregressive structure and need
not extend to designs in which a null covariate is nearly collinear with an
active one. Binary covariates are harder to detect: at $n = 50$ the active
covariate of the binary block is detected in $71.0\%$ of the replications,
against $93.7\%$ and $91.6\%$ for those of the correlated and independent
blocks, whereas the three rates exceed $97\%$ for $n \ge 100$. This is
expected, since a $\mathrm{Bernoulli}(0.5)$ covariate has half the standard
deviation of a standard normal one. As in the main design, the selected
Weibull fit may reach the exponential limit when the sample is small: this
happens in $49\%$ of the replications with $n = 50$, $16\%$ with $n = 100$ and
$0.6\%$ with $n = 200$, and its average sensitivity is then $0.734$, against
$0.989$ otherwise. The Gamma fit reaches its exponential limit in $2.6\%$ of
the replications, and every selected fit that did not converge, $0.5\%$ of the
$40{,}500$, is one of these boundary fits.

\section{Application}
\label{sec:application}

We illustrate the methodology on two datasets, a low-dimensional clinical
study and a gene expression study with $p > n$. In both, the three modal
families are fitted by Algorithm~\ref{alg:emlasso} and compared with the
EBIC-tuned log-Gaussian Lasso of Section~\ref{sec:simulation}. Since the model
is scale-equivariant, that is, replacing $y_i$ by $cy_i$ shifts $\gamma_0$ by
$\log c$ and leaves the slopes unchanged, the riboflavin response is
divided by its mean for numerical convenience. As in
Section~\ref{sec:simulation}, the paths of every method stop once more than
$\lfloor n/\log n \rfloor$ covariates are active.

\subsection{Survival after treatment for lung cancer}
\label{sec:app_lung}

The Veterans' Administration lung cancer trial \cite{kalbfleisch1980} records
the survival time in days of $137$ patients with advanced inoperable lung
cancer, together with nine covariates: treatment, Karnofsky performance score,
months from diagnosis, age, prior therapy and four cell-type indicators
(squamous, small, adeno and large). Only nine observations ($6.6\%$) are
right-censored and, since the present formulation does not accommodate
censoring, we analyze the $128$ complete cases, whose survival times are
strongly right-skewed, with a median of $62$ days, a maximum of $999$ and
sample skewness $2.98$. The results are reported in
Table~\ref{tab:app_lung}.

\begin{table}[!ht]
  \caption{VA lung cancer data, $n = 128$ complete cases, $p = 9$.
    Estimated coefficients at the EBIC-selected $\lambda$, where $\cdot$
    denotes a coefficient set to zero. The last column is a Weibull accelerated failure
    time (AFT) model with the four selected covariates, fitted to the $137$
    observations with censoring taken into account.}
  \label{tab:app_lung}
  \centering \small
  \begin{tabular}{l cccc c}
\toprule
& \multicolumn{3}{c}{Modal Lasso} & Log-Gaussian & Weibull AFT \\
\cmidrule(lr){2-4}
Covariate & Gamma & Weibull & Lognormal & Lasso & ($n=137$) \\
\midrule
Karnofsky score & $\phantom{-}0.027$ & $\phantom{-}0.027$ & $\phantom{-}0.026$ & $\phantom{-}0.026$ & $\phantom{-}0.029^{***}$ \\
Squamous & $\phantom{-}0.251$ & $\phantom{-}0.262$ & $\cdot$ & $\cdot$ & $\phantom{-}0.322$ \\
Small cell & $-0.253$ & $-0.241$ & $\cdot$ & $\cdot$ & $-0.386$ \\
Adeno & $-0.487$ & $-0.491$ & $\cdot$ & $\cdot$ & $-0.786^{**}$ \\
\midrule
$|\hat{\mathcal{S}}|$ & $4$ & $4$ & $1$ & $1$ & \\
$\hat\phi$ & $6.325$ & $0.039$ & $1.284$ & & \\
$\hat\lambda$ & $0.084$ & $0.084$ & $0.281$ & $0.281$ & \\
\bottomrule
\end{tabular}

  \vspace{2pt}
  {\footnotesize $^{**}p<0.01$, $^{***}p<0.001$. Treatment, months
  from diagnosis, age, prior therapy and the large-cell indicator were set to
  zero by every method; the large-cell indicator is collinear with the other
  three cell types.}
\end{table}

The Gamma and Weibull fits select the same four covariates, namely the
Karnofsky score and the squamous, small-cell and adeno indicators, with
similar coefficients and the expected signs: a higher Karnofsky score is
associated with longer survival, and the adeno and small-cell types with
shorter survival. As a check on the complete-case restriction, a Weibull AFT
model with these four covariates fitted to the $137$ observations, with
censoring taken into account, yields estimates of the same sign, larger in
magnitude as expected from the shrinkage of the Lasso, of which those of the
Karnofsky score and the adeno indicator are significant at the $1\%$ level.
The estimated shapes, $\hat{k}^{\ast} = 1.05$ for the Weibull and
$\hat\alpha = 1.16$ for the Gamma, are close to the exponential limit, at
which the mode tends to zero, so that the fitted modes are small relative to
the scale of the response.

The Lognormal fit retains only the Karnofsky score, and so does the
log-Gaussian Lasso, at the same $\hat\lambda = 0.281$ and with the same
coefficient, as stated in Proposition~\ref{prop:lognormal}. At this
$\lambda$ the two sets of slopes coincide and the intercepts differ by
$1.284$, which equals $\hat\phi$. Therefore, the cell-type effects identified by the Gamma
and Weibull families are not recovered after a logarithmic transformation of
the response.

\subsection{Riboflavin production in \emph{Bacillus subtilis}}
\label{sec:app_ribo}

The riboflavin dataset \cite{buhlmann2014} contains the logarithm of the
riboflavin production rate of \emph{B.~subtilis} and the expression levels of
$4088$ genes for $n = 71$ observations. We analyze the
\texttt{riboflavinV100} subset used by \cite{schelldorfer2011}, which
comprises the $p = 100$ genes with the largest empirical variances, and take
the production rate itself as the response, which is positive with sample
skewness $0.82$. The $71$ observations are repeated measurements on $28$
strains, with two to four observations per strain; as in
\cite{buhlmann2011} and \cite{buhlmann2014}, we treat them as independent. The
results at $\gamma_{\mathrm{EBIC}} = 0.5$ are reported in
Table~\ref{tab:app_ribo}.

\begin{table}[!ht]
  \caption{Riboflavin data, $n = 71$, $p = 100$
    (\texttt{riboflavinV100}), at $\gamma_{\mathrm{EBIC}} = 0.5$. Estimated
    coefficients on the log-mode scale for every gene selected by at least one
    method; $\cdot$ denotes a coefficient set to zero. YXLD\_at is the only
    gene of \cite{buhlmann2014} in this subset.}
  \label{tab:app_ribo}
  \centering \small
  \begin{tabular}{l cccc}
\toprule
& \multicolumn{3}{c}{Modal Lasso} & Log-Gaussian \\
\cmidrule(lr){2-4}
Gene & Gamma & Weibull & Lognormal & Lasso \\
\midrule
YCKE\_at & $\phantom{-}0.160$ & $\phantom{-}0.187$ & $\phantom{-}0.137$ & $\phantom{-}0.137$ \\
YCGN\_at & $-0.046$ & $-0.018$ & $-0.068$ & $-0.068$ \\
XHLA\_at & $\cdot$ & $\phantom{-}0.145$ & $\phantom{-}0.230$ & $\phantom{-}0.230$ \\
YXLE\_at & $-0.335$ & $-0.123$ & $\cdot$ & $\cdot$ \\
YTGB\_at & $-0.163$ & $-0.072$ & $\cdot$ & $\cdot$ \\
YXLD\_at & $\cdot$ & $\cdot$ & $-0.137$ & $-0.137$ \\
YCGM\_at & $-0.127$ & $-0.115$ & $\cdot$ & $\cdot$ \\
PCKA\_at & $\phantom{-}0.248$ & $\cdot$ & $\cdot$ & $\cdot$ \\
XLYA\_at & $\phantom{-}0.214$ & $\cdot$ & $\cdot$ & $\cdot$ \\
YXLJ\_at & $\cdot$ & $-0.116$ & $\cdot$ & $\cdot$ \\
ARGF\_at & $-0.113$ & $\cdot$ & $\cdot$ & $\cdot$ \\
ABH\_at & $\phantom{-}0.063$ & $\cdot$ & $\cdot$ & $\cdot$ \\
YCDH\_at & $-0.038$ & $\cdot$ & $\cdot$ & $\cdot$ \\
YVFL\_at & $\phantom{-}0.021$ & $\cdot$ & $\cdot$ & $\cdot$ \\
TRXA\_at & $\phantom{-}0.015$ & $\cdot$ & $\cdot$ & $\cdot$ \\
YHZA\_at & $-0.011$ & $\cdot$ & $\cdot$ & $\cdot$ \\
\midrule
$|\hat{\mathcal{S}}|$ & $13$ & $7$ & $4$ & $4$ \\
$\hat\phi$ & $0.107$ & $1.480$ & $0.390$ & \\
$\hat\lambda$ & $0.035$ & $0.158$ & $0.303$ & $0.303$ \\
\bottomrule
\end{tabular}
\end{table}

The Lognormal fit and the log-Gaussian Lasso select the same four genes,
with identical coefficients at the same $\hat\lambda = 0.303$; at this
$\lambda$ the slopes agree up to $4.4\times10^{-16}$ and the intercepts differ
by $0.390$, which equals $\hat\phi$, in agreement with
Proposition~\ref{prop:lognormal}. The Gamma and Weibull fits select
thirteen and seven genes, respectively; for the Gamma family the
criterion attains its minimum at the second-to-last point of the path, which
stops before more than $\lfloor 71/\log 71 \rfloor = 16$ genes become
active. YXLD\_at, the only one of
the three genes identified by \cite{buhlmann2014} that belongs to this subset,
is selected by the Lognormal fit, whereas the Gamma and Weibull fits select
YXLE\_at, whose sample correlation with YXLD\_at is $0.98$, so that the three
families respond to the same signal through different representatives.
Moreover, YCKE\_at and YCGN\_at are selected by the four methods.

\section{Concluding remarks}
\label{sec:conclusion}

In this paper, we have proposed a regularized version of parametric
modal regression for positive responses with high-dimensional covariates,
estimated by a proximal Newton algorithm. By writing
the density parameters of the Gamma, Weibull and Lognormal families as
functions of the conditional mode, each iteration reduces to a penalized
weighted least-squares problem with a pseudo-response and a working weight in
closed form, which is solved exactly by the coordinate-descent routines of
\texttt{glmnet}, so that the Lasso, adaptive Lasso and elastic-net penalties
are handled without further derivation. We have shown that the Lognormal
estimator coincides with the Gaussian Lasso applied to the log-transformed
response, that the conditional mean and the conditional mode share their
slopes under constant dispersion, and that the estimation error bound contains
no smoothing bias term.

The simulation studies show that the procedure detects the active
covariates once the sample size is moderate relative to the dimension, while
rarely selecting noise variables, with specificity above $0.94$ in every
design, and that it improves on the log-Gaussian Lasso for Gamma and, above
all, Weibull responses, with a lower estimation error in every configuration
of the main design. With weaker signals both methods lose power and the
comparison between them is unchanged, whereas $\gamma_{\mathrm{EBIC}} = 0$
increases the sensitivity substantially at a cost in specificity of at most
$0.031$. Autoregressive correlation among the
covariates does not degrade selection, and binary covariates require larger
samples than continuous ones. In the two applications, the Lognormal fit
reproduces the log-Gaussian Lasso, as established in
Proposition~\ref{prop:lognormal}, whereas the Gamma and Weibull fits identify
additional covariates, which in the lung cancer data are in agreement with a
censored Weibull regression fitted to the full sample.

The present formulation assumes complete and independent observations with a
constant dispersion, and it provides no standard errors for the selected
model, for which post-selection inference \cite{lee2016} is required. Natural
extensions include right-censored responses and random effects, where the
present surrogate would be combined with the E-step of the penalized EM
algorithms of Section~\ref{sec:pml}, and a dispersion that depends on the
covariates, under which the slopes of the conditional mean and of the
conditional mode no longer coincide. An in-depth investigation of such
extensions is beyond the scope of the present paper, but it constitutes an
interesting topic for further research.

\section*{Code and data availability}

Algorithm~\ref{alg:emlasso} is implemented in the \texttt{R} package
\texttt{ModalNet} \cite{modalnet}, version 0.3.1, available from the
Comprehensive R Archive Network (CRAN) and at
\url{https://github.com/chedgala/ModalNet}. The supplementary material
contains the scripts that reproduce every table of
Sections~\ref{sec:simulation} and~\ref{sec:application}, with replication $b$
generated under the seed $2025 + b$. The lung cancer data are distributed with
the \texttt{R} package \texttt{survival}, and the \texttt{riboflavinV100} data
with \texttt{ModalNet}.


\begin{thebibliography}{99}

\bibitem{buhlmann2011}
B\"uhlmann, P.\ and Van De Geer, S. (2011).
\textit{Statistics for High-Dimensional Data: Methods, Theory and Applications}.
Springer.

\bibitem{buhlmann2014}
B\"uhlmann, P., Kalisch, M., and Meier, L. (2014).
High-dimensional statistics with a view toward applications in biology.
\textit{Annual Review of Statistics and Its Application}, 1, 255--278.

\bibitem{kalbfleisch1980}
Kalbfleisch, J.~D.\ and Prentice, R.~L. (1980).
\textit{The Statistical Analysis of Failure Time Data}.
Wiley, New York.

\bibitem{chen2008}
Chen, J., and Chen, Z. (2008).
Extended Bayesian information criteria for model selection with large model spaces.
\textit{Biometrika}, 95(3), 759--771.

\bibitem{chen2016}
Chen, Y.-C., Genovese, C.~R., and Wasserman, L. (2016).
A comprehensive approach to mode clustering.
\textit{Electronic Journal of Statistics}, 10(1), 210--241.

\bibitem{chun2010}
Chun, H.\ and Kele\c{s}, S. (2010).
Sparse partial least squares regression for simultaneous dimension
reduction and variable selection.
\textit{Journal of the Royal Statistical Society: Series B}, 72(1), 3--25.

\bibitem{dempster1977}
Dempster, A.~P., Laird, N.~M., and Rubin, D.~B. (1977).
Maximum likelihood from incomplete data via the EM algorithm.
\textit{Journal of the Royal Statistical Society: Series B}, 39(1), 1--22.

\bibitem{freijeiro2022}
Freijeiro-Gonz\'alez, L., Febrero-Bande, M., and Gonz\'alez-Manteiga, W. (2022).
A critical review of LASSO and its derivatives for variable selection
under dependence among covariates.
\textit{International Statistical Review}, 90(1), 118--145.

\bibitem{friedman2010}
Friedman, J., Hastie, T., and Tibshirani, R. (2010).
Regularization paths for generalized linear models via coordinate descent.
\textit{Journal of Statistical Software}, 33(1), 1--22.

\bibitem{galarza2025modal}
Galarza, C.~E.\ and Lachos, V.~H. (2025).
Parametric modal regression for right-censored positive responses.
\textit{arXiv preprint}. Available at \url{https://github.com/chedgala/ModalCens}.

\bibitem{modalnet}
Galarza, C.~E., Lachos, V.~H., and Liu, D. (2026).
\textit{ModalNet: Regularized Parametric Modal Regression: Lasso, Elastic Net
and Ridge}. \texttt{R} package version 0.3.1.
\url{https://CRAN.R-project.org/package=ModalNet}.

\bibitem{green1990}
Green, P.~J. (1990).
On use of the EM algorithm for penalized likelihood estimation.
\textit{Journal of the Royal Statistical Society: Series B}, 52(3), 443--452.

\bibitem{lee2016}
Lee, J.~D., Sun, D.~L., Sun, Y., and Taylor, J.~E. (2016).
Exact post-selection inference, with application to the lasso.
\textit{Annals of Statistics}, 44(3), 907--927.

\bibitem{lee2014}
Lee, J.~D., Sun, Y., and Saunders, M.~A. (2014).
Proximal Newton-type methods for minimizing composite functions.
\textit{SIAM Journal on Optimization}, 24(3), 1420--1443.

\bibitem{loh2015}
Loh, P.-L.\ and Wainwright, M.~J. (2015).
Regularized $M$-estimators with nonconvexity: statistical and algorithmic
theory for local optima.
\textit{Journal of Machine Learning Research}, 16(19), 559--616.

\bibitem{mccullagh1989}
McCullagh, P.\ and Nelder, J.~A. (1989).
\textit{Generalized Linear Models} (2nd ed.).
Chapman \& Hall.

\bibitem{negahban2012}
Negahban, S.~N., Ravikumar, P., Wainwright, M.~J., and Yu, B. (2012).
A unified framework for high-dimensional analysis of $M$-estimators with
decomposable regularizers.
\textit{Statistical Science}, 27(4), 538--557.

\bibitem{oliveira2025}
Oliveira, D.~C.~R., Schumacher, F.~L., and Lachos, V.~H. (2025).
The use of the EM algorithm for regularization problems in
high-dimensional linear mixed-effects models.
\textit{Statistical Methods in Medical Research}. DOI: 10.1177/09622802251399913.

\bibitem{schelldorfer2011}
Schelldorfer, J., B\"uhlmann, P., and van de Geer, S. (2011).
Estimation for high-dimensional linear mixed-effects models using
$\ell_1$-penalization.
\textit{Scandinavian Journal of Statistics}, 38(2), 197--214.

\bibitem{tibshirani1996}
Tibshirani, R. (1996).
Regression shrinkage and selection via the lasso.
\textit{Journal of the Royal Statistical Society: Series B}, 58(1), 267--288.

\bibitem{wang2020robust}
Wang, Y., Liu, J., and Miao, B. (2020).
Robust variable selection and estimation based on kernel modal regression.
\textit{Entropy}, 22(9), 1043.

\bibitem{yao2014}
Yao, W.\ and Li, L. (2014).
A new regression model: modal linear regression.
\textit{Scandinavian Journal of Statistics}, 41(3), 656--671.

\bibitem{zhou2025}
Zhou, S., Yao, W., and others (2025).
Testing-driven variable selection in Bayesian modal regression.
\textit{arXiv preprint arXiv:2510.23831}.

\bibitem{zou2005}
Zou, H.\ and Hastie, T. (2005).
Regularization and variable selection via the elastic net.
\textit{Journal of the Royal Statistical Society: Series B}, 67(2), 301--320.

\bibitem{zou2006}
Zou, H. (2006).
The adaptive lasso and its oracle properties.
\textit{Journal of the American Statistical Association}, 101(476), 1418--1429.

\end{thebibliography}
\end{document}